\documentclass[aps,preprint,showpacs,groupedaddress]{revtex4}
\usepackage {graphicx,epsfig,graphics,color}

\begin{document}

\title{Localized ferromagnetic resonance force microscopy in permalloy-cobalt films}

\author{E.~Nazaretski}
\affiliation{Los Alamos National Laboratory, Los Alamos, NM
87545}
\author{Yu.~Obukhov}
\affiliation{Department of Physics, Ohio State University, Columbus
OH 43210,}
\author{I.~Martin} \affiliation{Los Alamos National
Laboratory, Los Alamos, NM 87545}
\author{D.~V.~Pelekhov}
\affiliation{Department of Physics,
Ohio State University, Columbus OH 43210,}
\author{K.~C.~Cha}
\affiliation{Los Alamos National Laboratory, Los Alamos, NM 87545}
\author{E.~A.~Akhadov}
\affiliation{Los Alamos National Laboratory, Los Alamos, NM 87545}
\author{P.~C.~Hammel}
\affiliation{Department of Physics, Ohio State University, Columbus OH 43210,}
\author{R.~Movshovich}
\affiliation{Los Alamos National Laboratory, Los Alamos, NM
87545}


\begin{abstract}
We report Ferromagnetic Resonance Force Microscopy (FMRFM)
experiments on a justaposed continuous films of permalloy and
cobalt. Our studies demonstrate the capability of FMRFM to perform
local spectroscopy of different ferromagnetic materials. Theoretical
analysis of the uniform resonance mode near the edge of the film
agrees quantitatively with experimental data. Our experiments
demonstrate the micron scale lateral resolution in determining local
magnetic properties in continuous ferromagnetic samples.

\end{abstract}

 \maketitle

Magnetic resonance force microscopy (MRFM) is attracting increasing
attention as a result of its high spin sensitivity and excellent
spatial resolution in paramagnetic and nuclear spin
systems.\cite{Rugar 1992, Rugar 1994, Zhang 1996a,Rugar 2004,Degen
2005, Rugar 2009} MRFM studies on microfabricated and continuous
ferromagnetic samples have been also performed. \cite{Zhang 1996,
Loubens 2007, Nazaretski 2006a,LocalFMR:obukhov:prl:2008, Nazaretski
2009} Here we report FMRFM experiments performed on a
non-overlapping permalloy (Py) and cobalt (Co) continuous films and
demonstrate the capability of FMRFM to spectroscopically identify
the distinct magnetic properties of two adjacent ferromagnetic
films. We quantitatively model the resulting force signal strength
and compare it with the experimental data.



\begin{figure}
\includegraphics [angle=0,width=8.5cm]{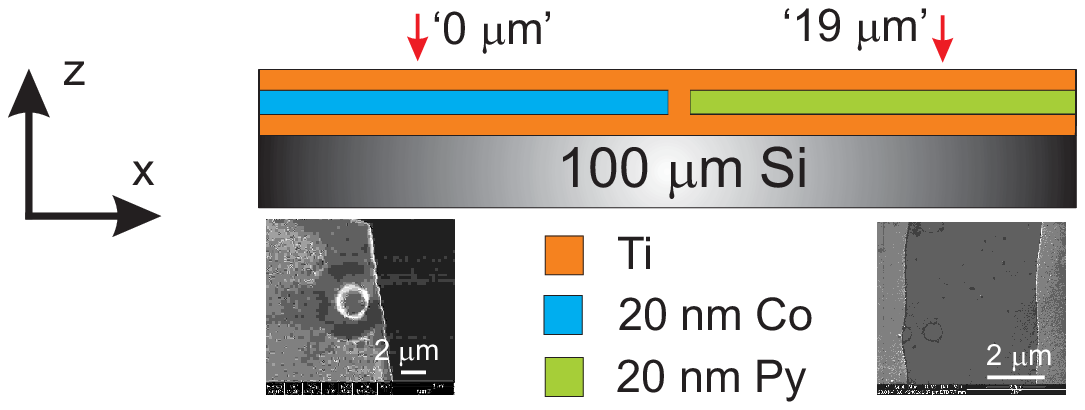}
\caption{} \label{sample}
\end{figure}

The permalloy-cobalt sample is schematically shown in Fig.
\ref{sample}. A 20 nm thick Ti film was uniformly applied onto the
surface of a 100 $\mu$m thick Si (100) wafer. 20 nm of Co was
deposited into a rectangular area (2.5 $\times$ 5 mm) defined in
photoresist followed by the lift-off. A complimentary rectangular
area of 20 nm thick Py was similarly defined and deposited. The
entire structure was then coated with a 20 nm thick layer of Ti. The
interface between the Co and Py regions was examined in a scanning
electron microscope (SEM) and revealed a gap whose width varies
between 3 and 6 $\mu$m along the entire length of the sample (see
SEM image in Fig. \ref{sample}). An approximately 1.7 $\times$ 1.7
mm$^2$ piece was cut and glued to the stripline resonator of the
FMRFM apparatus and the film plane was oriented perpendicular to the
direction of the external magnetic field $H_{\rm ext}$. For FMRFM
studies we used the cantilever with the spherical magnetic tip (see
SEM image in Fig. \ref{sample}) and its spatial field profile has
been carefully characterized \cite{Nazaretski 2008}. More
details on the experimental apparatus can be found in Ref. \cite{Nazaretski 2006}\\

\begin{figure}
\includegraphics [width=8.5cm]{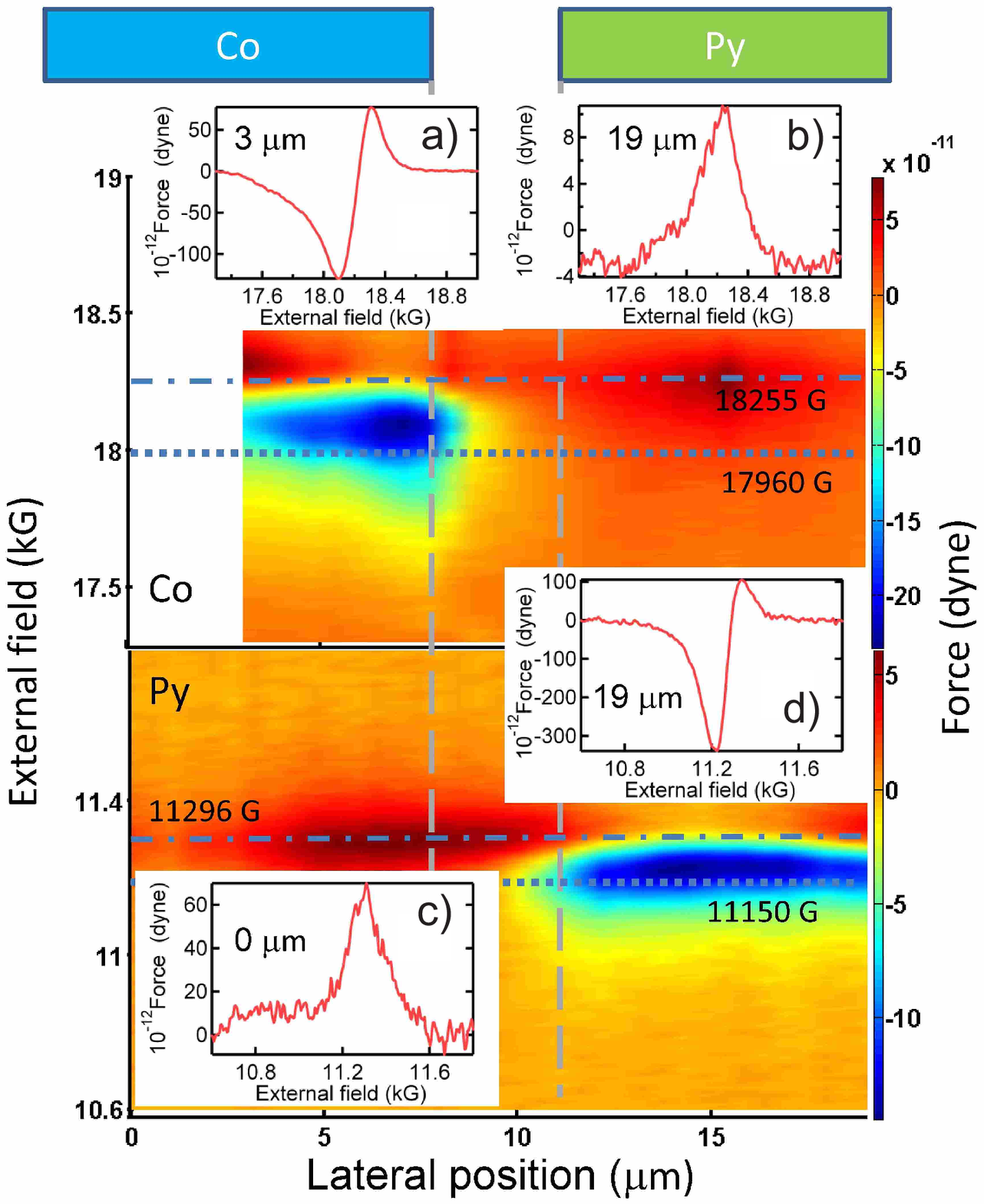}
\caption{} \label{Figure 1}
\end{figure}

In Fig.~\ref{Figure 1} we show the evolution of the FMRFM signal as
a function of the lateral position and applied magnetic field. The
cantilever was scanned across the interface between Co and Py, in
the region indicated by arrows in Fig. \ref{sample}. The FMRFM
signal was recorded in two different regions of $H_{ext}$ which
correspond to Py and Co resonance fields for the microwave frequency
of $f_{RF}$=9.35 GHz. Insets in Fig.~\ref{Figure 1} show the
evolution of the FMRFM spectra as a function of lateral position.
The signal, reminiscent of those reported earlier in
\cite{Nazaretski 2007}, is comprised of two distinctive
contributions. The first, a negative signal which occurs at lower
values of $H_{ext}$ is a localized resonance originating from the
region of the sample right under the cantilever tip where the probe
field is strong and positive. The second contribution is positive
and is observed at higher values of $H_{ext}$. This signal arises
from a larger region of sample remote from the tip which, therefore,
experience a weak negative tip field; we will label this the
``uniform resonance''. As seen in Fig.~\ref{Figure 1}, at the
beginning of the lateral scan the negative (lower field) resonance
structure is present only in the Co spectrum (see inset a)). Near a
lateral position of 9 $\mu$m we see no localized signals (with
negatively shifted $H_{ext}$) for either the Py or the Co signals.
However upon scanning further over the Py film, the Py resonance
begins to show a localized signal, while the Co signal continues to
show only a uniform (positively shifted $H_{ext}$) signal (inset
b)).


We analyze the uniform contribution to the FMRFM signals considering
the case when the entire dynamic magnetization $m$ is constant and
the resonance field is only weakly affected by the probe. This
approximation is valid for the large probe-sample distances (insets
b) and c) in Fig. \ref{Figure 1}).   The frequency of the uniform
resonance in a thin
 film can be written as  $\omega_{RF}/\gamma=H_{ext}-4\pi
 M_s$, where $4\pi M_s$ is the saturation magnetization and $\gamma$ is the gyromagnetic ratio.
FMRFM spectra shown in b) and c) insets in Fig. \ref{Figure 1} yield
the values of $4\pi M_s$ = 8052 G for Py and $4\pi M_s$ = 15013 G
for Co respectively.

For quantitative analysis of the FMRFM data it is important to have
an accurate estimate of the probe-sample separation. Magnetic Force
Microscopy (MFM) measurements were used to calibrate the
probe-sample separation. The cantilever was scanned across the Py -
Co interface and changes in its resonance frequency were recorded.
The gradient of the MFM force for a semi-infinite film can be
written as follows:
\begin{equation}
\frac{\partial F}{\partial
z}=4m_pM_sL\frac{x(x^2-3z^2)}{(x^2+z^2)^3},
\label{Eq:MFM_force_gardient}
\end{equation}
where $m_p$ = 7$\times$10$^{-9}$ emu is the probe magnetic moment
\cite{Nazaretski 2008} and $L$ is the film thickness. $z$ is the
probe-film distance and $x$ is the lateral position with respect to
the film edge ($x\geq 0$). MFM data were acquired at H$_{ext}$ =
18255 G, thus, both films were saturated. The MFM data and the fit
to Eq.~\ref{Eq:MFM_force_gardient} are shown in Fig.~\ref{Figure
2}a, yielding the tip-sample separation $z$ $\approx$ 4.4 $\mu$m and
the films boundaries ($x$ $\leq$ 8 $\mu$m for Co and $x$ $\geq$ 11
$\mu$m for Py).

\begin{figure}
\includegraphics [width=8.5cm]{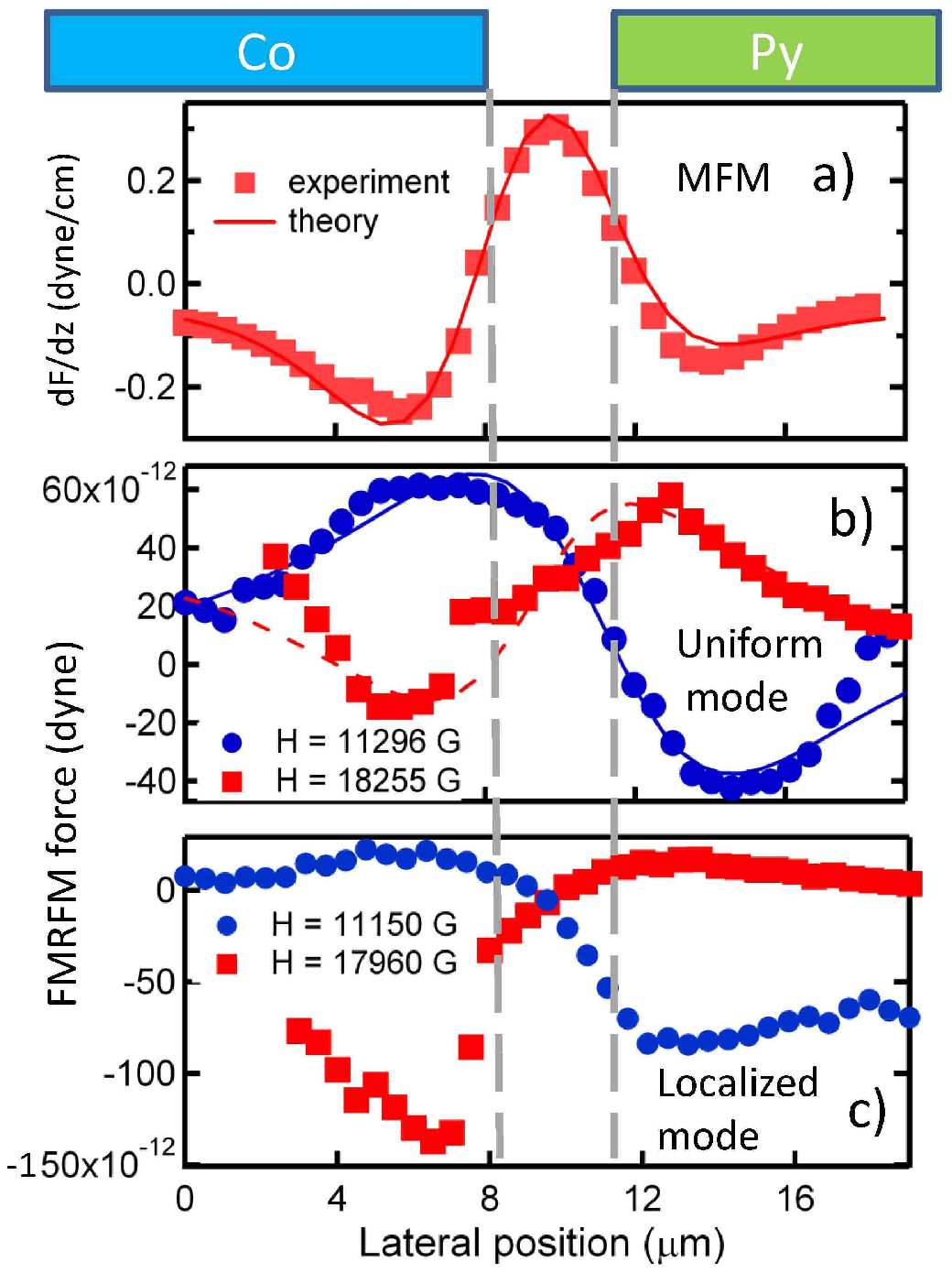}
\caption{} \label{Figure 2}
\end{figure}

The tip field suppresses the uniform FMR mode in the region under
the tip, and according to Obukhov {\it et al.} \cite{Obukhov:film
2008} the magnitude of the suppression depends on the tip-sample
separation. It is described as partial suppression at distances
$z\gg\sqrt{\frac{2m_p}{\pi M_sL\alpha_0}}$ ($\alpha_0$ is the first
zero of the Bessel function $J_0(\alpha_0)=0$) and full suppression
at  $z\ll\sqrt{\frac{2m_p}{\pi M_sL\alpha_0}}$. The region of
suppressed magnetization is confined to a region of radius
$r=\sqrt{2}z$. FMRFM data discussed here were taken at the boundary
of these two regions, thus we consider the regime of full
suppression, however we introduce the magnitude of the suppression
as a fit parameter. Ferromagnetic resonance excitation generates a
precessing transverse magnetization $m$, thus reducing $M_z$; the
change of $M_z$ is $\delta M_z=\sqrt{M_s^2-m^2}-M_s \approx
-m^2/2M_s$. Here we modulate the amplitude of $m$ with a 100$\%$
modulation depth at the cantilever resonance frequency. The FMRFM
force exerted on a cantilever is $F=-\int Lm^2/2M_s\cdot\partial
H_p/\partial z dr'$, where integration is performed over the entire
film area. The total FMRFM force close the edge of the film is well
approximated by
\begin{equation}
F=-\frac{m^2}{2M_s}L\left(\frac{4xzm_p}{(x^2+z^2)^2}-\beta\int_S\theta(x')\frac{\partial H_p}{\partial z}
                    (x-x')dr'\right),
                    \label{Eq:FMRFM uniform}
\end{equation}
where the first term describes the force between the probe and the
semi-infinite film and the second term represents the force between
the probe and the area of radius $r=\sqrt{2}z$ under the tip. The
Heaviside function $\theta(x')$ represents the fact that the film is
positioned at $x'$ $\geq$ 0 and the dimensionless parameter $\beta$
quantifies the degree of suppression of the uniform FMR mode. In
Fig.~\ref{Figure 2}b we plot the experimental data extracted from
Fig. \ref{Figure 1} and corresponding fits using Eq. \ref{Eq:FMRFM
uniform}.  Fig. \ref{Figure 2}b demonstrates good qualitative and
quantitative agreement between theory and experiment and
demonstrates the validity of the model. It is important to mention
that in our model we assume the dynamic magnetization $m$ to be
constant throughout the film. However, $m$ may vary due to the
change of the demagnetizing field e.g. $-4\pi M_s$ far from the film
boundary and $-2\pi M_s$ at the film boundary. Our estimates show
that $m$ changes from a constant value in the film down to zero at
the film edge.  The length scale of this change is $\pi M_sL/\Delta
H$ $\approx$ 1 $\mu$m ($\Delta H$ is the linewidth of the uniform
resonance), small compared to the probe-sample distance thus only
weakly affecting the fits shown in Fig.~\ref{Figure 2}b.

The spatial resolution of the uniform FMR mode shown in
Fig.~\ref{Figure 2}b is comparable to the MFM lateral resolution
depicted in Fig.~\ref{Figure 2}a and is determined by the
probe-sample separation of  z $\approx$ 4.4 $\mu$m. However, it can
be further improved by tracking the intensity of the FMRFM signal at
values of H$_{ext}$ lower than that of the uniform FMR mode (insets
a) and d) in Fig. \ref{Figure 1}). In Fig.~\ref{Figure 2}c we show
the FMRFM force acquired at H$_{ext}$ = 17960 G for Co and H$_{ext}$
= 11150 G for Py respectively (values of H$_{ext}$ are schematically
indicated with dotted lines in Fig.~\ref{Figure 1}). The
contribution to the FMRFM signal at lower values of H$_{ext}$
originates from the localized region of the sample under the probe.
As seen in Fig.~\ref{Figure 2}c the lateral resolution is on the
order of 3 $\mu$m (10$\%$ - 90$\%$ change in localized signal
intensity) and is determined by the FMR resonance linewidth and the
spatial profile of the FMR mode under the tip. Further theoretical
and numerical analysis is required to understand the evolution of
the FMR modes excited under the probe in the presence of a strongly
inhomogeneous tip field and
boundaries of the sample.\\


In conclusion, we have obtained local FMR spectra in justaposed
ferromagnetic samples and our quantitative model for the spatial
variation of the uniform mode agrees well with experimental data. We
have demonstrated spectroscopic imaging of Py and Co semi-infinite
films with the spatial resolution for the tip induced
resonance of $\approx$ 3 $\mu$m.\\


The work performed at Los Alamos was supported by the US Department
of Energy, and Center for Integrated Nanotechnologies at Los Alamos
and Sandia National Laboratories. The work at OSU was supported by
the US Department of Energy through grant DE-FG02-03ER46054.

\newpage

Figure Caption

\vspace{3cm}

Figure 1: Schematic of the Co - Py sample. The arrows mark the scan
range for spectra shown in Fig. \ref{Figure 1}. The SEM image on the
right shows the gap between Py and Co. The SEM image on the left
depicts the cantilever tip.

\vspace{2cm}

Figure 2: FMRFM force image as a function of $H_{ext}$ and the
lateral position. We show the Co and Py forces in the upper and
lower panels respectively. Insets a) - d) demonstrate the evolution
of the FMRFM signal as a function of lateral position indicated on
the left-hand side of each inset. Vertical dashed lines show the
boundaries of the Co and Py films. The horizontal dashed-doted lines
are drawn through the values of H$_{ext}$ = 18255 G for Co and
H$_{ext}$ = 11296 G for Py respectively and correspond to the
uniform resonance (see Fig.~\ref{Figure 2}b). The horizontal
  dotted lines at H$_{ext}$ = 17960 G and H$_{ext}$ = 11150 G for Co
  and Py respectively, mark the localized FMRFM signals.
Experimental parameters: T = 11 K, $f_{RF}$=9.35 GHz, probe-sample
distance $\approx$ 5.6 $\mu$m.

\vspace{2cm}

Figure 3: a) MFM data acquired at H$_{ext}$=18255 G, solid line is
the fit to Eq.~\ref{Eq:MFM_force_gardient}. b) FMRFM force data for
the uniform ferromagnetic resonance (FMR) modes. H$_{ext}$ = 18255 G
for Co (squares) and H$_{ext}$ = 11296 G for Py (circles). Solid and
dashed lines are fits of Eq.~\ref{Eq:FMRFM uniform} to the data. Fit
parameters: $m/M_s$ = 0.0014, $\beta$ = 0.65 for Co and  $m/M_s$ =
0.0028, $\beta$ = 0.5 for Py. c) FMRFM force for the localized
(close to the probe) FMR mode acquired at H$_{ext}$ = 17960 G for Co
(squares) and H$_{ext}$ = 11150 G for Py (circles).  The lateral
resolution is better than 3 $\mu$m.

\newpage

\newpage

\newpage

\newpage

\begin{thebibliography}{99}

\bibitem{Rugar 1992}
D. Rugar, C. S. Yannoni, and J. A. Sidles, Nature {\bf 360}, 563
(1993)

\bibitem{Rugar 1994}
D. Rugar and O. Z\"uger and S. Hoen and C. S. Yannoni and H. M.
Vieth and R. D. Kendrick, Science {\bf 264}, 1560 (1994)

\bibitem{Zhang 1996a}
Z. Zhang, M. L. Roukes, and P. C. Hammel, J. Appl. Phys. {\bf 80},
6931 (1996)

\bibitem{Rugar 2004}
D. Rugar, R. Budakian, H. J. Mamin,  and W. Chui, Nature  {\bf 430},
329 (2004)

\bibitem{Degen 2005}
C. L. Degen, Q. Lin, A. Hunkeler, U. Meier, M. Tomaselli, and B. H.
Meier, Phys. Rev. Lett. {\bf 94}, 207601 (2005)

\bibitem{Rugar 2009}
C. L. Degen, M. Poggio, H. J. Mamin, C. T. Rettner, and D. Rugar,
Proc. Natl. Acad. Sci. {\bf 106}, 1313 (2009)

\bibitem{Zhang 1996}
Z. Zhang, P. C. Hammel, and P. E. Wigen, Appl. Phys. Lett. {\bf 68},
2005 (1996)

\bibitem{Loubens 2007}
G. de Loubens, V.V. Naletov, O. Klein, J. Ben Youssef, F. Boust, and
N. Vukadinovic, Phys. Rev. Lett. {\bf 98}, 127601 (2007)


\bibitem{LocalFMR:obukhov:prl:2008}
 Yu.~Obukhov,  D.~V.~Pelekhov,  J.~Kim,  P.~Banerjee,  I.~Martin,  E. Nazaretski, R.~Movshovich,
 S.~An,  T.~J.~Gramila, S.~Batra,  P.~C.~Hammel,
  Phys. Rev. Lett. {\bf 100}, 197601, (2008)

\bibitem{Nazaretski 2009}
E. Nazaretski, D. V. Pelekhov, I. Martin, M. Zalalutdinov, D.
Ponarin, A. Smirnov, P. C. Hammel, and R. Movshovich, Phys. Rev. B
{\bf 79}, 132401 (2009)

\bibitem{Nazaretski 2006a}
E. Nazaretski, J. D. Thompson, M. Zalalutdinov, J. W. Baldwin, B.
Houston, T. Mewes, D. V. Pelekhov, P. Wigen, P. C. Hammel,  and R.
Movshovich, J. Appl. Phys. {\bf 101}, 074905 (2007)



\bibitem{Nazaretski 2008}
E. Nazaretski, E. A. Akhadov, I. Martin, D. V. Pelekhov, P. C.
Hammel, and R. Movshovich, Appl. Phys. Lett. {\bf 92}, 214104
(2008).

\bibitem{Nazaretski 2006}
E. Nazaretski, T. Mewes, D. Pelekhov, P. C. Hammel, and R.
Movshovich, AIP Conf. Proc. {\bf 850}, 1641 (2006)

\bibitem{Nazaretski 2007}
E. Nazaretski, D. V. Pelekhov, I. Martin,  M. Zalalutdinov, J. W.
Baldwin, T. Mewes, B. Houston, P. C. Hammel, and R. Movshovich,
Appl. Phys. Lett. {\bf 90} 234105 (2007)

\bibitem{Obukhov:film 2008}
 Yu.~Obukhov, D.~V.~Pelekhov, E.~Nazaretski,
R.~Movshovich, P.~C.~Hammel, Appl. Phys. Lett. {\bf 94}, 172508
(2008)



\end{thebibliography}
\end{document}